\documentclass[aps,prb,superscriptaddress,twocolumn,nopacs,amsmath,amssymb,letter,graphicx]{revtex4-1}

\usepackage{color}
\usepackage{graphicx}
\usepackage{dcolumn} 
\usepackage{bm}
\usepackage{xspace}
\setcitestyle{numbers,square}

\begin{document}

\title{Dual quantum confinement and anisotropic spin splitting\\ in the multi-valley semimetal PtSe$_2$}

\author{O.~J.~Clark}
\affiliation {SUPA, School of Physics and Astronomy, University of St. Andrews, St. Andrews KY16 9SS, United Kingdom}

\author{F.~Mazzola}
\affiliation {SUPA, School of Physics and Astronomy, University of St. Andrews, St. Andrews KY16 9SS, United Kingdom}

\author{J.~Feng}
\affiliation {SUPA, School of Physics and Astronomy, University of St. Andrews, St. Andrews KY16 9SS, United Kingdom}
\affiliation {Suzhou Institute of Nano-Tech. and Nanobionics (SINANO), CAS, 398 Ruoshui Road, SEID, SIP, Suzhou, 215123, China}

\author{V. Sunko}
\affiliation {SUPA, School of Physics and Astronomy, University of St. Andrews, St. Andrews KY16 9SS, United Kingdom}
\affiliation {Max Planck Institute for Chemical Physics of Solids, N{\"o}thnitzer Stra{\ss}e 40, 01187 Dresden, Germany}

\author{I. Markovi\'c}
\affiliation {SUPA, School of Physics and Astronomy, University of St. Andrews, St. Andrews KY16 9SS, United Kingdom}
\affiliation {Max Planck Institute for Chemical Physics of Solids, N{\"o}thnitzer Stra{\ss}e 40, 01187 Dresden, Germany}

\author{L. Bawden}
\affiliation {SUPA, School of Physics and Astronomy, University of St. Andrews, St. Andrews KY16 9SS, United Kingdom}

\author{T.~K.~Kim}
\affiliation{Diamond Light Source, Harwell Campus, Didcot, OX11 0DE, United Kingdom}

\author{P.~D.~C.~King}
\email{philip.king@st-andrews.ac.uk}
\affiliation {SUPA, School of Physics and Astronomy, University of St. Andrews, St. Andrews KY16 9SS, United Kingdom}

\author{M.~S.~Bahramy}
\email{bahramy@ap.t.u-tokyo.ac.jp}
\affiliation{Quantum-Phase Electronics Center and Department of Applied Physics, The University of Tokyo, Tokyo 113-8656, Japan}
\affiliation{RIKEN center for Emergent Matter Science (CEMS), Wako 351-0198, Japan} 

\date{\today}

\begin{abstract}
{
\noindent We investigate the electronic structure of a two-dimensional electron gas created at the surface of the multi-valley semimetal 1T-PtSe$_2$. Using angle-resolved photoemission and first-principles-based surface space charge calculations, we show how the induced quantum well subband states form multiple Fermi surfaces which exhibit highly anisotropic Rashba-like spin splittings. We further show how the presence of both electron- and hole-like bulk carriers causes the near-surface band bending potential to develop an unusual non-monotonic form, with spatially-segregated electron accumulation and hole accumulation regions, which in turn amplifies the induced spin splitting. Our results thus demonstrate the novel environment that semimetals provide for tailoring electrostatically-induced potential profiles and their corresponding quantum subband states. 
}
\end{abstract}

\maketitle    

\section{Introduction}
Gate-voltage control of surface and interface charge carrier densities not only lies at the heart of modern microelectronic devices such as the ubiquitous semiconductor transistor~\cite{ando_electronic_1982}, but also provides a mechanism for stabilising new physical regimes. When a two-dimensional electron gas (2DEG) is created via electrical gating, for example, the associated breaking of inversion symmetry can lead to a lifting of the spin-degeneracy of its electronic states via the so-called Rashba effect~\cite{bychov_oscillatory_1984}. This is a striking result of spin-orbit coupling in solids, and can lead to new approaches for utilising the electron spin in so-called spintronics~\cite{koga_rashba_2002, koo_control_2009, datta_electronic_1990}. The injection of large sheet carrier densities, utilizing approaches such as ionic liquid gating~\cite{goldman_electrostatic_2014,yuan_high_2009}, has further established the potential to obtain gate-voltage control over the collective states of materials, for example inducing superconducting instabilities~\cite{ueno_electric_2008,ye_superconducting_2012,ueno_discovery_2011} or manipulating magnetic order~\cite{yamada_electrically_2011, yang_anisotropic_2015,wei_prediction_2017, wen_evolution_2018}. As such, modifying surface space-charge regions has become a powerful route to control and explore the electronic properties of materials, both for applications and to facilitate fundamental understanding.

This approach has been widely applied to the semiconducting transition-metal dichalcogenides (TMDs), part of a diverse materials family~\cite{chhowalla_chemistry_2013} which has been attracting significant recent attention both in bulk and single-layer form~\cite{chhowalla_chemistry_2013, rossnagal_origin_2011, manzeli_2D_2017, xu_spin_2014, bahramy_ubiquitous_2018}. Electrical gating has been employed to realise monolayer field-effect transistors~\cite{mak_valley_2014, gong_electric_2014, amani_electrical_2013}, to create tunable Zeeman-like spin splittings~\cite{riley_negative_2015, yuan_zeeman_2013}, to manipulate charge order~\cite{li_controlling_2016}, and to induce unconventional superconducting states~\cite{lu_evidence_2015, saito_superconductivity_2016,li_controlling_2016}. Here, we investigate a TMD, $1T$-PtSe$_2$, whose ground state is not a semiconductor, but a semimetal. It has a metallic temperature-dependent resistivity~\cite{yang_quantum_2018}, and as such would not typically be considered as a natural system for electrical gating. We nonetheless show that an electrostatic band bending potential can be created to screen a surface (or equivalently gate-induced) charge, which in turn confines a well-defined 2DEG. We further show that the band-bending potential has an unconventional non-monotonic form arising due to an intrinsic competition between the electron and hole-like carriers in screening the surface charge, and that the induced 2DEG exhibits a pronounced anisotropic Rashba-like spin splitting. Our work therefore provides new insight into the potential for the all-electrical control of spin-polarised charge carriers in semimetallic systems, and into the unconventional space-charge regions that can be created in multi-valley and multi-band systems.

\section{Methods}
Angle-resolved photoemission spectroscopy (ARPES) was performed at the I05 beamline of Diamond Light Source~\cite{hoesch_facility_2017}, using linear-horizontal ($p$) polarised light with photon energies between 20 and 120~eV. Single crystal samples of PtSe$_2$ were mounted using conventional methods (conductive silver epoxy on to a sample plate) and cleaved \textit{in-situ} via the top-post method at the measurement temperature ($<14$~K). Rb was deposited on the cleaved surface, again at the measurement temperature, using a well-outgassed SAES Rb getter source operated at 5.6~A. From analysis of integrated spectral weight of Rb~4p, Pt~4f and Se~3d core-levels as measured by x-ray photoemission spectroscopy ($h\nu=150$~eV), as well as the lack of any free-electron gas states observed in ARPES associated with formation of monolayer Rb islands~\cite{Eknapakul}, we conclude that the coverage was sub-monolayer. 

Relativistic density-functional theory (DFT) calculations  were performed using the Perdew-Burke-Ernzerhof exchange-correlation functional corrected by the semilocal Tran-Blaha-modified Becke-Johnson potential, as implemented in the~{\sc wien2k} package~\cite{wien}. The Brillouin zone (BZ) was sampled by a $20\times 20\times 20$ $k$-mesh and  the muffin-tin radius $R_{\text{MT}}$ for all atoms was chosen such that its product with the maximum modulus of reciprocal vectors $K_{\text{max}}$ becomes $R_{\text{MT}}K_{\text{max}}$ = 7.0. To describe the surface electronic structure, the bulk DFT calculations were downfolded using maximally localized Wannier functions~\cite{souza, mostofi, kunes} purely made of Se-$p$ orbitals, and the resulting 12-band tight-binding transfer integrals implemented within a 100-unit cell supercell with an additional on-site potential term representing the electrostatic band bending potential. This was solved self-consistently with Poisson's equation, assuming a static bulk permittivity $\varepsilon=\kappa \varepsilon_0$ with $\varepsilon_0$ being the vacuum permittivity and $\kappa$ being the dielectric constant fixed at 40~\cite{lei_2017}. The only adjustable parameter is the total amount of band bending, varied until the experimentally-observed surface sheet carrier density $N_\mathrm{ss} \approx 1\times10^{14}$~cm$^{-2}$ is achieved. 

\section{Results and Discussions}
 \begin{figure}[t!]
 \includegraphics[width=\columnwidth]{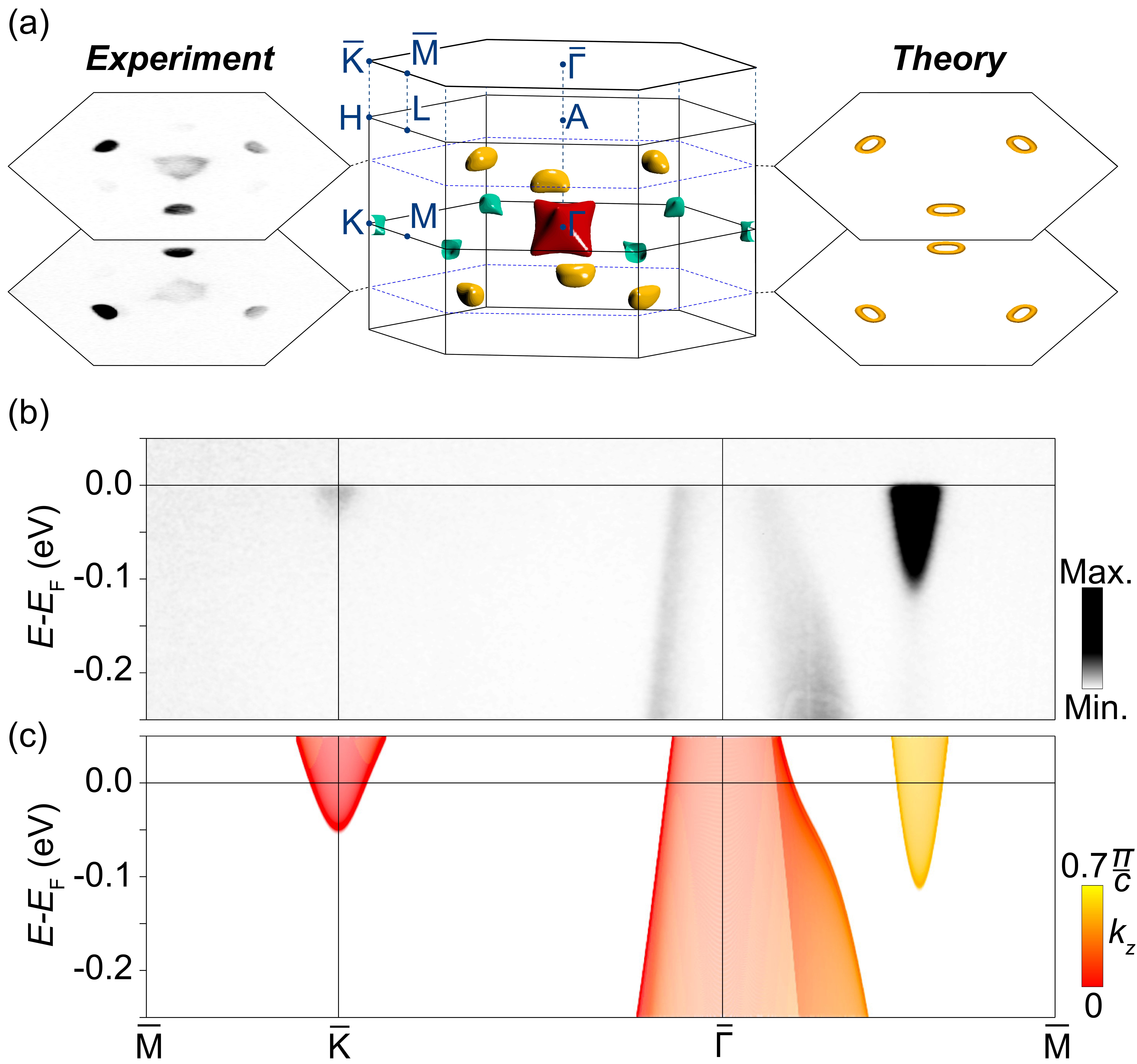}
 \caption{(a) Three-dimensional Fermi surface of bulk PtSe$_2$, extracted from DFT calculations. Additional $k_x$-$k_y$ contours are displayed for $k_z=\pm 0.6 \frac{\pi}{c}$ as extracted from photon energy dependent ARPES, left (h$\nu$=120~eV and 99~eV, $E_F \pm$ 30~meV) and from DFT, right. (b) ARPES dispersions along the $\overline{\text{M}}$-$\overline{\text{K}}$-$\overline{\Gamma}$-$\overline{\text{M}}$ path (h$\nu$=53~eV).
 (c) Equivalent dispersions extracted from DFT calculations, projected as a function of $k_z$ between 0 and $0.7\frac{\pi}{c}$ (see colour bar), which are in good agreement with the experimental measurments. No additional states are present in this energy range for other $k_z$ values.\label{f:fig1} }
 \end{figure}

Figure~\ref{f:fig1} shows the bulk three-dimensional Fermi surface and near-Fermi level ($E_{\mathrm{F}}$) band dispersions of PtSe$_2$, as obtained from both DFT calculations and photon energy-dependent ARPES measurements. Three sets of Fermi pockets are visible, all of predominantly Se $p$-orbital character due to the non-bonding nature of Pt in PtSe$_2$ \cite{chhowalla_chemistry_2013, guo_electronic_1986}. A hole pocket is located at the $\Gamma$ point in the BZ centre. This is the upper part of a type-II Dirac cone that is formed by the crossing of Se $p_z$ and $p_{x/y}$ states below the Fermi level~\cite{bahramy_ubiquitous_2018, huang_type_2016, yan_lorentz_2017, zhang_experimental_2017}. An electron pocket is located in the $k_z=0$ plane at the K points of the BZ. Six additional electron pockets are  centered at low-symmetry $k$-points within the $\Gamma$-M-L-A planes, hereafter referred to as S-pockets. Three of these pockets are located at positive $k_z$, with a three-fold in-plane distribution reflecting the trigonal symmetry of the crystal structure; the Kramers doublet partners of these electron pockets are located at negative $k_z$, with their in-plane momenta rotated by 180 degrees with respect to the upper pockets (Fig.~\ref{f:fig1}(a)). 
The volume of the electron and hole pockets is equal, thereby making PtSe$_2$  a compensated semimetal with a complex multi-valley Fermi surface. From our DFT calculations, we estimate the corresponding bulk electron and hole concentrations to be $(n,p)\approx{1.7}\times10^{20}$~cm$^{-3}$, equivalent to  about 0.01 electrons per formula unit. The majority of the electron count is contributed by the S-pockets ($\approx{1.2}\times10^{20}$~cm$^{-3}$) with only a small contribution from the electron pockets at K. 

\begin{figure*}[ht]
\includegraphics[width=\textwidth]{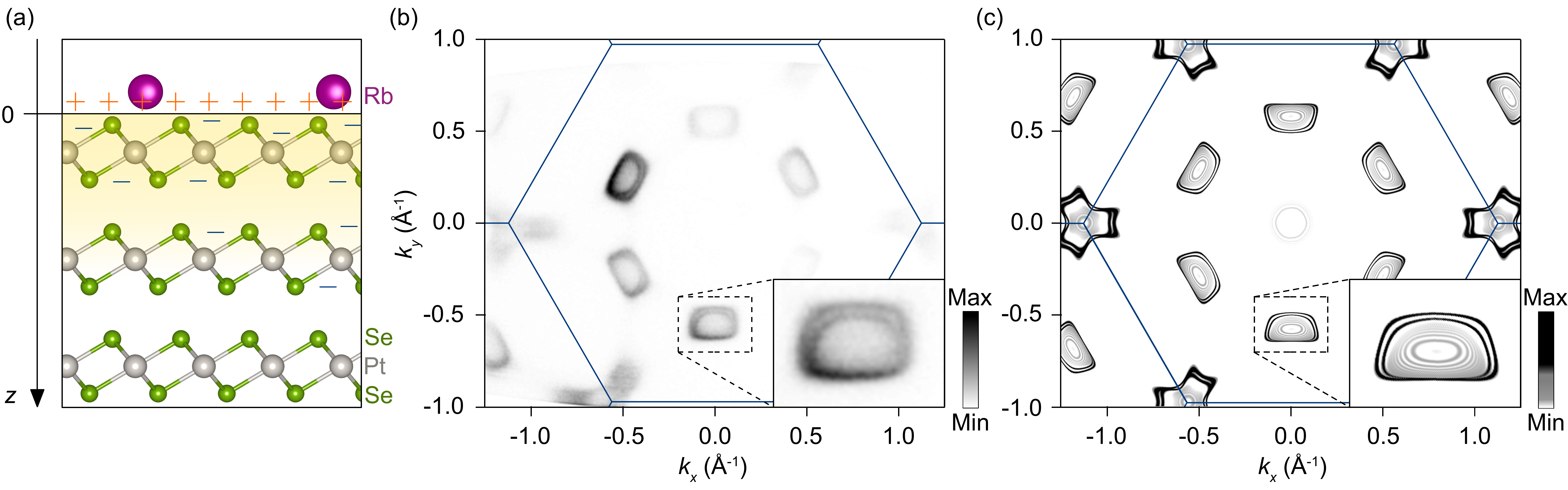}
\caption{ \label{f:fig2} (a) Schematic representation of near-surface charge accumulation when Rb atoms are deposited on a surface. (b) Measured Fermi surface of Rb-dosed PtSe$_2$ ($E_F \pm 12$~meV, $h\nu$=37~eV). (c) Equivalent Fermi surface resulting from a self-consistent band bending calculation.}
\end{figure*}

We show below how this electronic structure becomes strongly modified at the surface following the adsorption of small quantities of Rb (Fig.~\ref{f:fig2}). Rb is highly electropositive, and so readily donates electrons into the near-surface region, akin to injecting carriers in a field-effect style device (Fig.~\ref{f:fig2}(a))~\cite{riley_negative_2015}. We find that this causes the formation of large propeller-like Fermi surfaces at the zone-corners, $\overline{\text{K}}$-points, and a pair of near-rectangular states, yielding a sharp and intense rim of spectral weight around each of the $\overline{\Gamma}$-$\overline{\text{M}}$ centered bulk  S-pockets (Fig.~\ref{f:fig2}(b)). No band folding is observed, ruling out surface reconstruction driven by the surface Rb absorption. Rather, we attribute all of these new states as quantum well states of the original bulk bands, now confined by a near-surface band bending potential.

Consistent with this, our photon energy-dependent measurements (Supplementary Fig.~S1~\cite{Supp}) show that the new surface-related bands are two-dimensional, unlike the underlying bulk S pockets~\cite{clark_fermiology_2018}. We have therefore created a well-defined two-dimensional electron gas at the surface of PtSe$_2$. This might seem surprising given the metallic conductivity of the bulk. However, as discussed above, PtSe$_2$ is not a true metal, but a semimetal with small electron and hole pockets. Given their small Fermi wavevectors $k_\mathrm{F}$, the corresponding Thomas-Fermi screening length $L_{\mathrm{TF}}\sim\!1/k_\mathrm{F}$ should be comparable to a moderately-doped semiconductor, unlike for a true metal where this becomes extremely small. Moreover, the surface sheet density extracted from the Luttinger area of the new surface Fermi pockets observed in Fig.~\ref{f:fig2}(b) is $N_\mathrm{ss} \approx 1\times10^{14}$~cm$^{-2}$. This represents approximately an order of magnitude larger charge carrier accumulation as compared to the carrier density of the bulk, and so can trigger a pronounced surface band bending to ensure charge neutrality. 

To further verify that the above considerations enable PtSe$_2$ to support a 2DEG defined by a near-surface band bending, we have performed self-consistent surface space-charge calculations via solution of a coupled Poisson-Schr\"odinger equation implemented into a realistic tight-binding model (see methods described above). In this way, we incorporate the realistic multi-band bulk electronic structure as determined from advanced {\it ab initio} calculations, benchmarked against our experimental measurements in Fig.~\ref{f:fig1}, and incorporate the effects of charge carrier screening through a standard dielectric picture. We further neglect additional energy shifts arising from exchange and correlation terms, which are known to lead to negative electronic compressibility for low-density 2DEGs induced in other TMDs such as WSe$_2$~\cite{riley_negative_2015}. This is justified by the relatively high net electron densities contributed by the bulk (given the semi-metallic nature) and the 2DEG. The total energy of the resulting 2DEG is, thus, expected to be dominated by the kinetic term arising from the dynamics of these existing and accumulated carriers.

\begin{figure}[t]
	\includegraphics[width=\columnwidth]{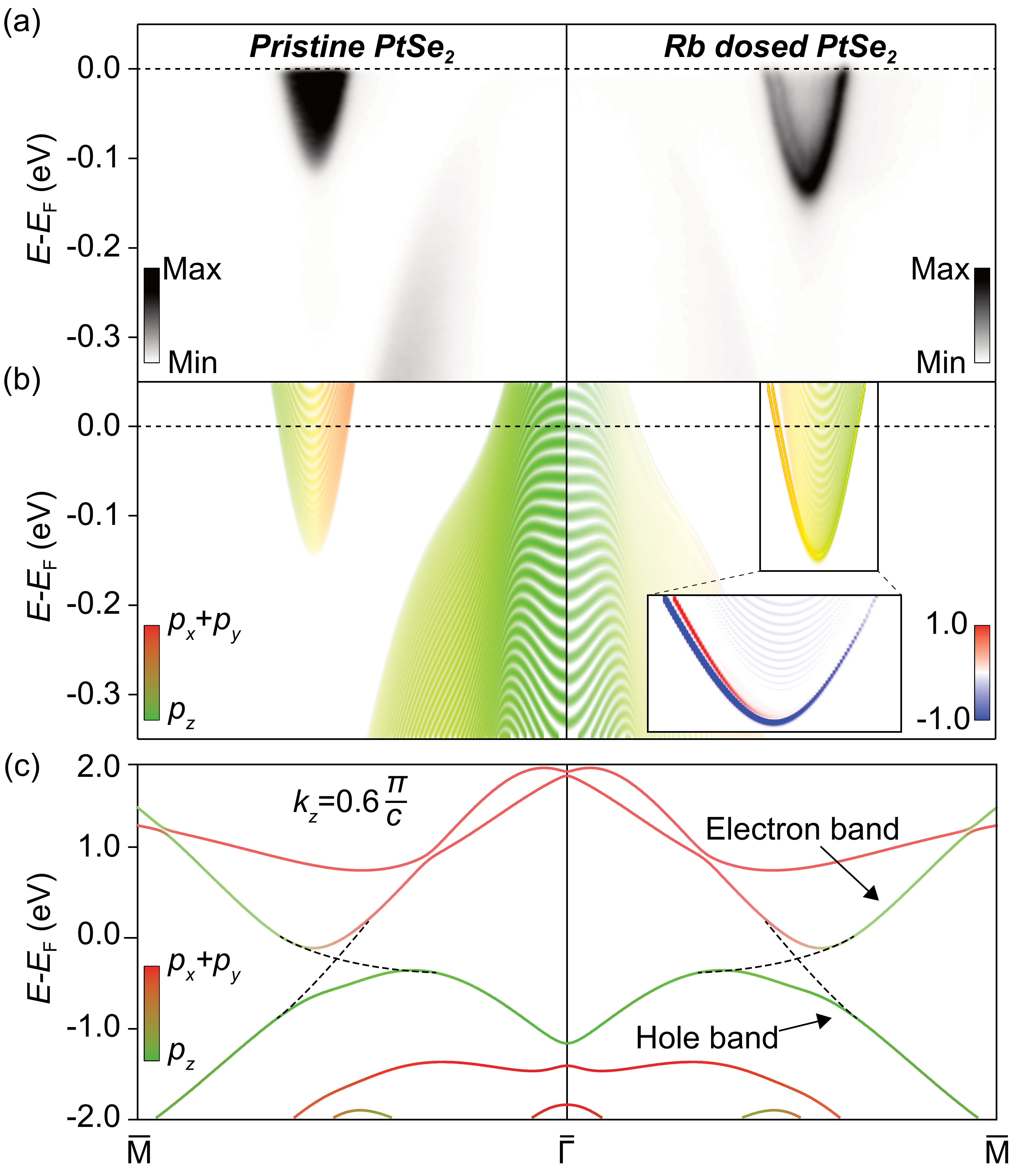}
	\caption{ \label{f:fig3}  (a) Pristine (left) and Rb-dosed (right) ARPES dispersions ($h\nu=37$~eV) measured along the $\overline{\Gamma}-\overline{\text{M}}$ direction. (b) Orbitally-projected surface slab calculations over an equivalent range as for (a). The inset shows the spin polarisation of the confined Rashba-split pair along $\overline{\Gamma}-\overline{\text{M}}$, projected onto the perpendicular (i.e., chiral) spin component. (c) Bulk orbitally-projected band structure of pristine PtSe$_2$ calculated along the in-plane $\overline{\Gamma}$-$\overline{\text{M}}$ direction for $k_z=$0.6$\frac{\pi}{c}$. The dashed lines show schematically the expected band structure in the absence of hybridisation between the $p_{x,y}$ and $p_z$ orbitals.} 
\end{figure}

The resulting surface Fermi surface (Fig.~\ref{f:fig2}(b,c)) and band dispersions (Fig.~\ref{f:fig3}(a,b)) are  in excellent agreement between our calculations and experimental measurements. The slab calculations assume an ideal bulk-truncated surface, with no structural modifications or distortions. The good agreement between our calculations and experimental measurements indicates that the dominant influence of the surface Rb absorption is therefore a donation of electrons into the near-surface region, and the results found here from both experiment and theory can be considered purely as arising from electrostatic doping effects.

Excitingly, projecting our calculated dispersion along $\overline{\Gamma}$-$\overline{\text{M}}$ onto the spin component perpendicular to this direction (Fig.~\ref{f:fig3}(b) inset) allows us to identify the band splitting evident in both our calculations (Fig.~\ref{f:fig2}(c) and Fig.~\ref{f:fig3}(b)) and experiment (Fig.~\ref{f:fig2}(b) and Fig.~\ref{f:fig3}(a)) as a spin splitting of oppositely polarised states. Such a spin-splitting is generically allowed when inversion symmetry is broken, whereby spin-orbit interactions may then lift the spin degeneracy. Our calculations show that the spin lies within the surface plane along the $\overline{\Gamma}$-$\overline{\text{M}}$ direction, pointing perpendicular to the in-plane momentum. Such spin-momentum locking is indicative of an out-of-plane potential gradient as the symmetry-breaking potential here, leading to a spin-splitting of the Rashba-type~\cite{bychov_oscillatory_1984}. 

We thus attribute the observed spin splittings as arising due to the presence of the near-surface band bending potential. In some 2H-TMDs, such a bending potential can effectively reveal a pre-existing intrinsic spin-polarisation of different energy valleys by lifting the layer degeneracy of their wave functions strongly localised on the topmost and adjacent layers~\cite{riley_negative_2015,yuan_zeeman_2013,Riley_NP,Kim_NL}. The observed spin-polarisation is thus merely out-of-plane and isotropic, being set by a hidden in-plane dipolar field locally existing within each layer of the bulk crystal structure due to their lack of inversion symmetry when considered individually~\cite{Riley_NP,Zhang_NP}. Here, however, the conduction band wave functions are highly delocalized, and so the spin polarisation is predominantly induced by the gradient of the bending potential, $\nabla V$, via a $k$-dependent planner spin-orbital field $(\nabla V \times {\bf k}) $, and the spin polarisation manifests as a Rashba-like spin texture. 

The spin-splitting is, however, highly anisotropic. As evident in both our experiment and calculations, the spin splitting is only clearly resolved on the low-momentum side of the electron pocket, persisting down to the band minimum, but being strongly suppressed on the high-momentum side of this S pocket (Figs.~\ref{f:fig2} and \ref{f:fig3}). Our orbital-projected calculations (Fig.~\ref{f:fig3}(c)) reveal how such an anisotropic spin splitting of the surface 2DEG is a natural consequence of the mixed orbital character of the underlying bulk electron pocket. This pocket is formed from the hybridisation of a $p_z$ and $p_{x/y}$ band as they disperse in-plane. The resulting hybridised band therefore has predominantly $p_z$ orbital character on its outer edge and a much more mixed $p_z$--$p_{x/y}$ orbital character on its inner edge (Fig.~\ref{f:fig3}(c)). This orbital character switch across the pocket is inherited by the 2DEG state when confined by the band bending potential (Fig.~\ref{f:fig3}(b)). On the high momentum, predominately $p_z$-derived, side the orbital angular momentum (OAM) is small, and so a negligible spin splitting develops (i.e., $\bf{L}\cdot\bf{S}$ is small). On the other hand, where the orbital character becomes strongly mixed, a large OAM can be expected and so the spin splitting becomes much larger~\cite{park_orbital_2011, kim_nature_2014,sunko_maximal_2017}.  Thus the multi-orbital character drives a highly anisotropic spin splitting to develop in this system. Similar effects are also evident for the $\overline{\text{K}}$-point pockets, with a maximal spin splitting along the $\overline{\Gamma}-\overline{\text{K}}$ direction, (Fig.~\ref{f:fig2}(c)).

While anisotropic, the maximum spin splitting achieved here is, in fact, rather large, with a momentum splitting at the Fermi level along $\overline{\Gamma}-\overline{\text{M}}$ of $\Delta{k}_{F}=0.025 \pm 0.001$~\AA$^{-1}$. This at least an order of magnitude larger than what is typically achieved in gated semiconductor 2DEGs such as InAs~\cite{koo_control_2009}. This is in part due to the relatively large spin-orbit coupling strength of Se $4p$ orbitals. More than that, however, it indicates that inversion symmetry must be broken strongly within the 2DEG here~\cite{sunko_maximal_2017}. 

\begin{figure}[t]
	\includegraphics[width=\columnwidth]{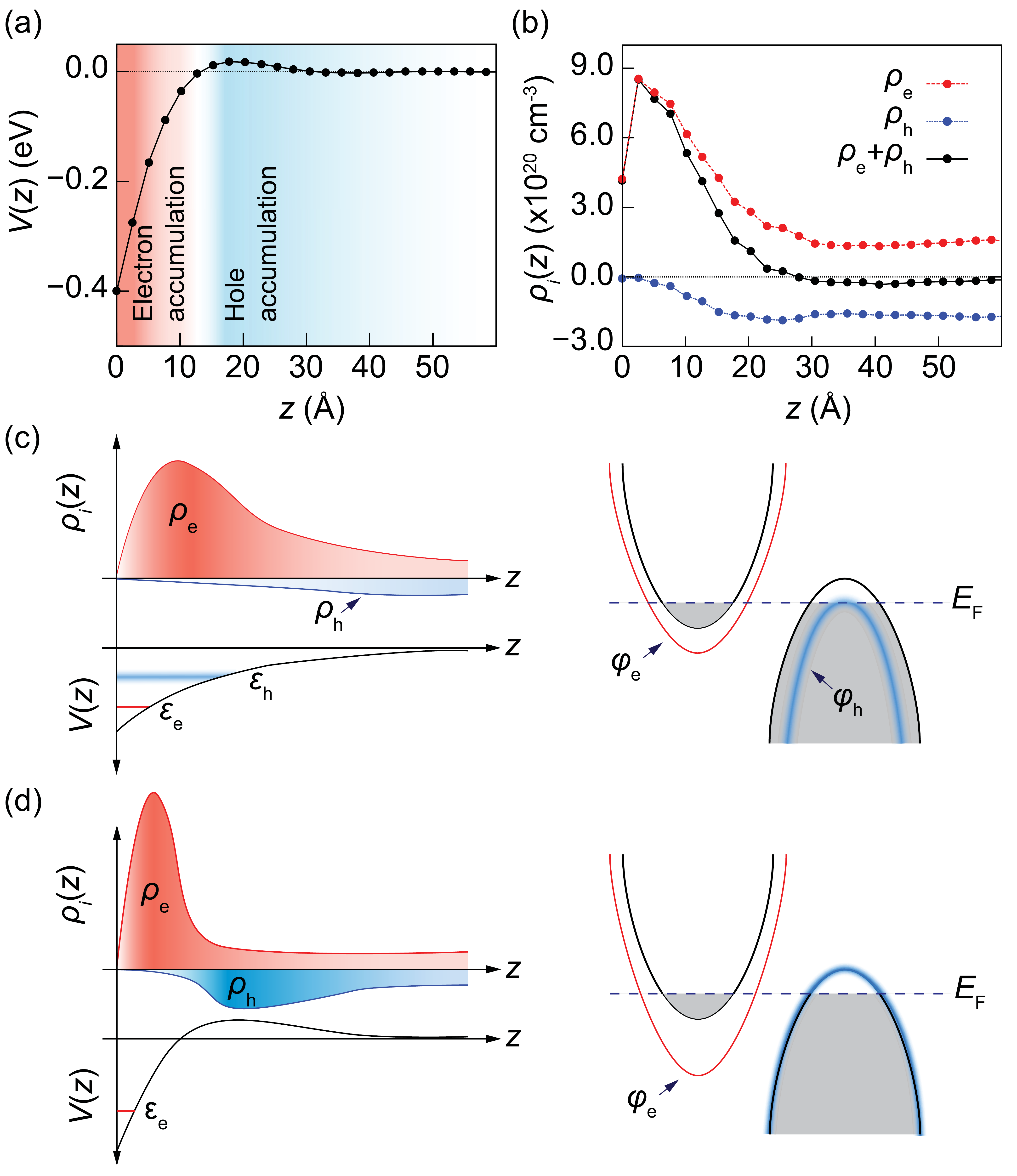}
	\caption{ \label{f:fig4} Calculated (a) potential profile and (b) charge density profile for surface-doped PtSe$_2$ with a surface sheet carrier density $N_\mathrm{ss} \approx 1\times10^{14}$~cm$^{-2}$. In (b), $\rho_h$ and $\rho_e$ denote the individual contribution of the hole and electron bands to the charge density, respectively. (c, d) Schematic illustration of quantum confinement of surface states in a compensated semimetal, considering (c) a conventional band bending potential and (d) a real potential satisfying the Poisson equation for such a double-carrier system.}
\end{figure}
To probe this, we investigate the near-surface potential profile, $V(z)$, resulting from our self-consistent band bending calculations. Surprisingly, we find that this exhibits an unconventional form, exhibiting a sign change in  $V(z)$  approximately $15$~\AA~below the surface (Fig.~\ref{f:fig4}(a)). This ``dual'' band bending would not be expected in a conventional space-charge region: The potential profile is given via solution of Poisson's equation, $\nabla^2V(z)=-\rho(z)/\varepsilon$ where $\rho(z)=e[\rho_e(z)-\rho_{e,b}-\rho_h(z)+\rho_{h,b}]$ is the depth-dependent space charge, $\{e,h\}$ denote electron and hole, and $b$ denotes the bulk, subject to the boundary conditions $\nabla{V(z)}|_{z=0}=eN_\mathrm{ss}/\varepsilon$ and $V(z)\rightarrow0$ as $z\rightarrow\infty$. A positive surface charge dictates that $V(0)$ must be negative. For a single carrier type, solution of Poisson's equation enforces the potential to always have a negative curvature,  implying that the potential profile varies monotonically with depth below the surface, asymptotically approaching zero towards the bulk over a length scale of approximately the Thomas-Fermi screening length~\cite{moench_semiconductor}. 

PtSe$_2$ is, however, not a single-carrier system, but, as discussed above, a compensated semimetal with a Fermi surface composed of both electron and hole-like carriers. As shown in Fig.~\ref{f:fig4}(c), a conventional downward band bending is able to effectively confine the the wave functions of electron-like carriers in such a system near to the surface (indicated as $\varphi_e$), leading to a well-defined 2DEG, $\varepsilon_e$. On the other hand, the surface hole-like bands are shifted downwards in energy such that their wave functions ($\varphi_h$) strongly overlap with the bulk continuum. The accumulated surface electrons of the hole pocket (here the $\Gamma$-pocket) will therefore be delocalized far away from the surface, over a length scale significantly exceeding the Thomas-Fermi screening length. 

\begin{figure}[t]
\includegraphics[width=\columnwidth]{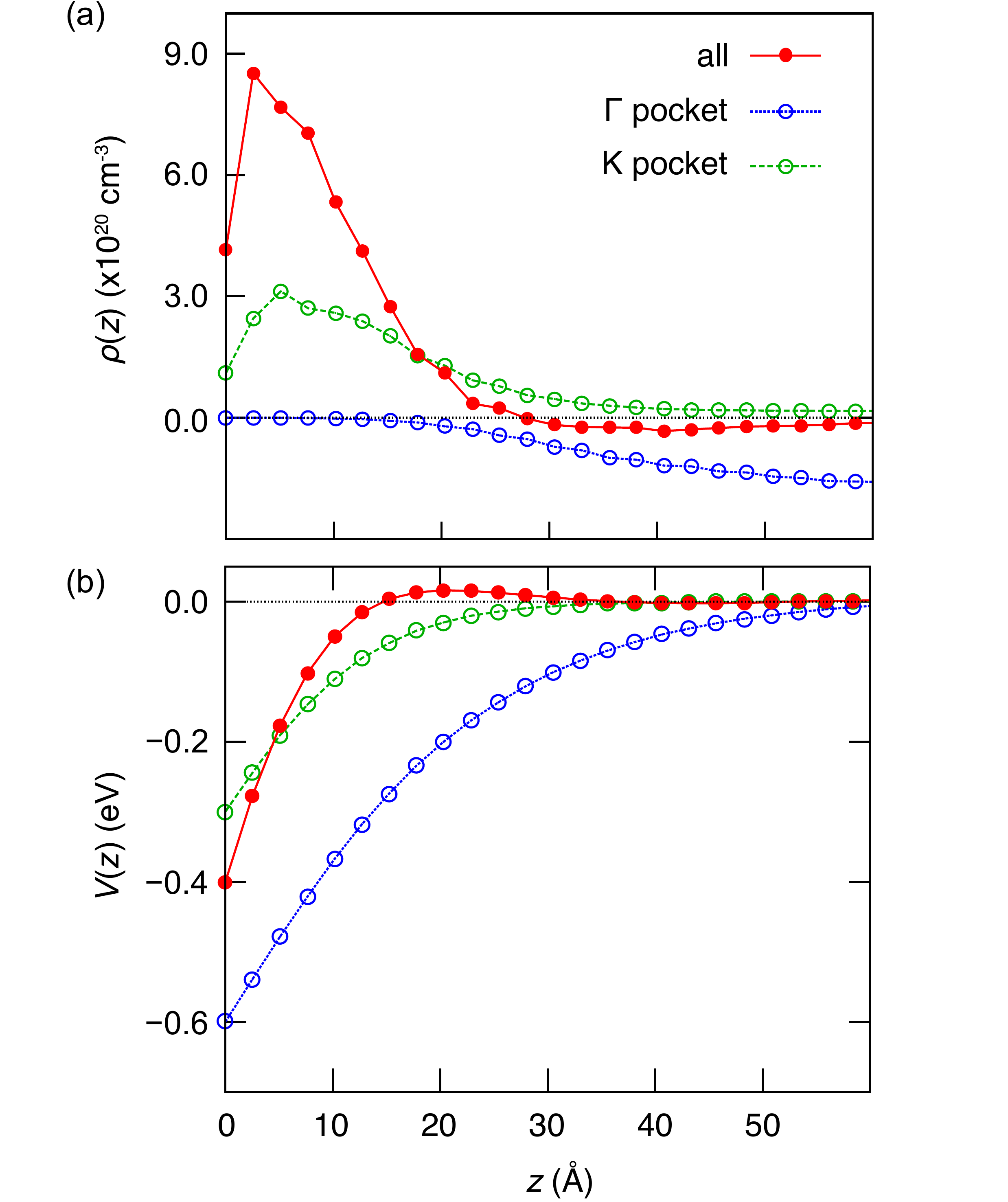}
\caption{ \label{f:v_rho} (a) Comparison of  calculated charge density profile resulting from all Fermi pockets of PtSe$_2$ vs. a hypothetical case where only the electron (K-) pocket and hole ($\Gamma$-) pockets are included in the space-charge calculations. (b) The corresponding potential profile for each case. For all three cases, the total sheet carrier density is fixed at $10^{14}$ cm$^{-2}$.   }  
\end{figure}

To avoid this energetically unfavourable condition, we find that a positive potential barrier is formed beneath the surface, as shown from explicit calculations in Fig.~\ref{f:fig4}(a), and represented schematically in Fig.~\ref{f:fig4}(d). The main role of this potential barrier is to minimize the overlap between the surface and bulk hole states. In this way,  charge is effectively pumped between the electron and hole states, allowing the surface charge to still be screened rapidly, confining the electrons close to the surface, but leaving the holes much closer to a charge-neutral configuration as a function of depth. The result is a surface electron accumulation region (required due to the positive surface charge, implying $\nabla{V(z)}|_{z=0}> 0$ and, hence, $V(0)<0$) which is separated from the bulk region by an intermediate hole accumulation region ($V(z)>0$, Fig.~\ref{f:fig4}(a,b)). Thus, a well-defined 2DEG is still developed, while avoiding the penalty of large carrier delocalisation lengths of the hole gas associated with the same band bending potential.

\begin{figure}[h]
\includegraphics[width=\columnwidth]{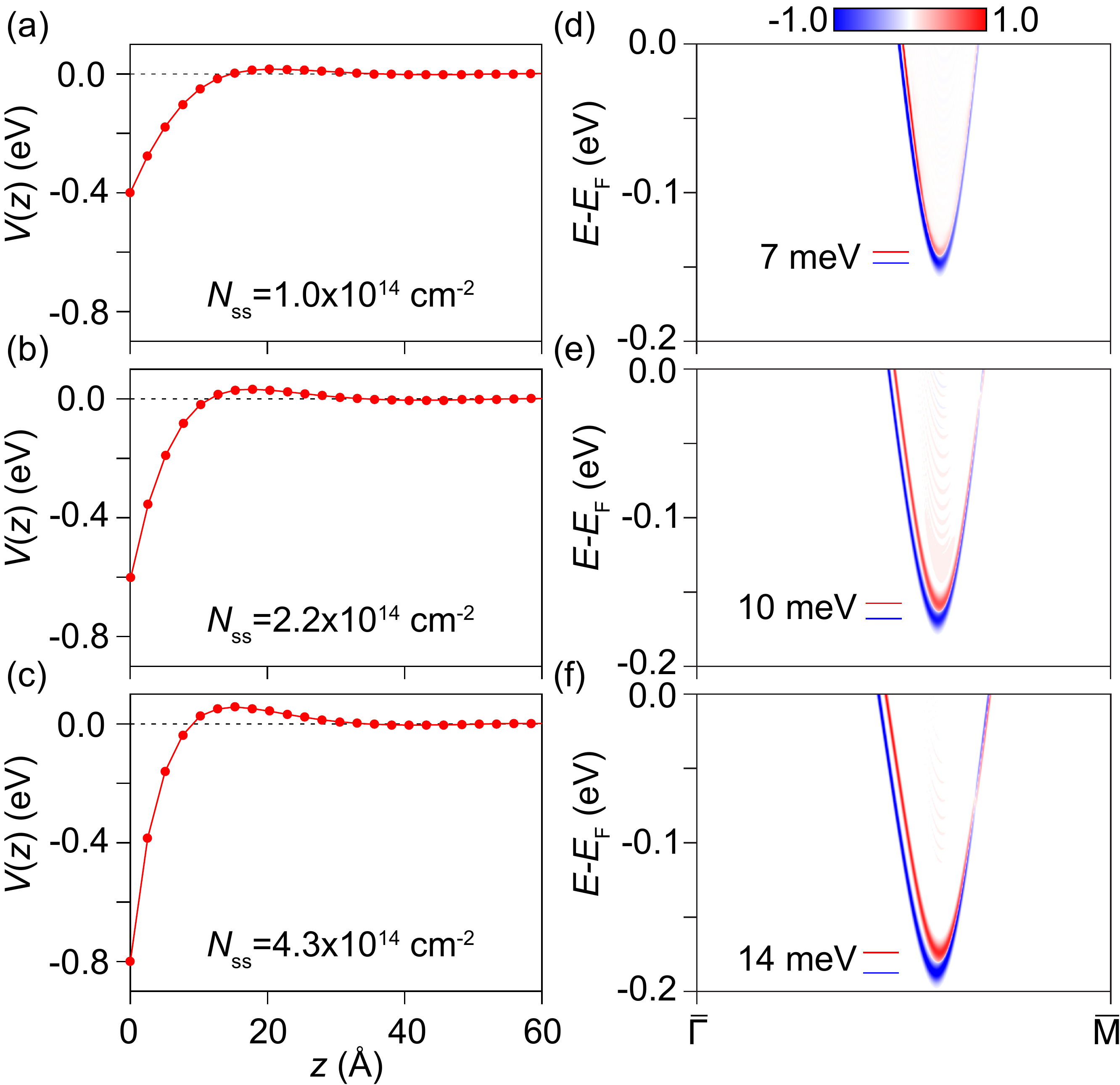}
\caption{ \label{f:bb} Expected  potential profile of surface-doped PtSe$_2$ at sheet carrier densities (a) $1\times10^{14}$ cm$^{-2}$, (b) $2.2\times10^{14}$ cm$^{-2}$, and (c) $4.3\times10^{14}$ cm$^{-2}$. (d-f) The respective spin-resolved surface electronic structures along the $\overline{\Gamma}-\overline{\text{M}}$ direction. Here, spin projection direction is set to be normal to both the momentum direction and the gradient of the bending potential. The size of the spin splitting grows with increasing magnitude of the near-surface potential gradient, indicating that the asymmetric band bending potential is the dominant driver of the Rashba-like spin splitting here.} 
\end{figure}

We note that the band bending potential observed here is, therefore, a general consequence of the multi-carrier nature of PtSe$_2$, and thus should be a general feature of systems with both electron- and hole-like bulk carriers. Indeed, repeating our calculations for PtSe$_2$ but including only the electron-like (K) or hole-like ($\Gamma$) pockets in isolation, we find a conventional monotonic form of the near-surface band bending (See Fig.~\ref{f:v_rho}). This confirms the above conclusions that the  multi-carrier nature is critical in generating the non-monotonic near-surface potential profiles. Intriguingly, comparing the full calculation to that with only pockets of a single carrier type included indicates that the combination of both electron and hole pockets causes the total potential profile to become much steeper near the surface (Fig.~\ref{f:v_rho}). This steeper band bending potential is key to enhancing the energy scale of inversion symmetry breaking, helping to stabilise the large spin splittings observed here and discussed above. To further support this picture, we show in Fig.~\ref{f:bb} the resulting spin-split states for three potential profiles applied to PtSe$_2$, accumulating different sheet carrier densities $N_{ss}$. As can be seen by enhancing the gradient of potential at the surface (i.e. increasing $N_{ss}$), the spin splitting of the quantised subband of the S-pocket undergoes a monotonic increase. 

\section{Conclusions}
Our findings show that a 2DEG can be created in the semimetal PtSe$_2$. Moreover, they demonstrate that semimetals in general provide an unusual surface space-charge environment as a result of a competition between electron- and hole-like carriers. Together with spin-orbit coupling, we have observed how this can magnify Rashba-like spin splittings. More generally, the resulting potential, which is confining for both holes and electrons in spatially-separated regions, could provide a natural potential barrier to screen  surface-induced charges from the underlying semimetallic bulk. Beyond improved spintronic functionality, enhanced potential gradients could also provide new routes to optimise charge carrier separation between bulk and surface regions and to controllably manipulate the relative population of electron- and hole-like carriers, driving a cross-over from multi-carrier to single-carrier type transport at the surface - a new form of rectification. Together, our findings thus motivate the further study and potential exploitation of the often overlooked semimetal for a range of possible device functionalities and as an environment for stabilising novel physics. 

\

\begin{acknowledgements}
We thank M. D. Watson for useful discussions. We gratefully acknowledge support from the Leverhulme Trust (Grant No.~RL-2016-006), the Royal Society, the European Research Council (Grant No. ERC-714193-QUESTDO) CREST, JST (No. JPMJCR16F1), and the International Max-Planck Partnership for Measurement and Observation at the Quantum Limit. OJC, VS, and LB acknowledge EPSRC for PhD studentship support through grant Nos. EP/K503162/1, EP/L015110/1 and EP/G03673X/1. IM acknowledges PhD studentship support from the IMPRS for the Chemistry and Physics of Quantum Materials. We thank Diamond Light Source (via Proposal Nos. SI13438-1, SI16262-2 and SI18705-1) for access to the I05 beamline that contributed to the results presented here. The research data supporting this publication can be accessed at Ref.~\cite{metadata}.
\end{acknowledgements}


\begin{thebibliography}{30}

\bibitem{ando_electronic_1982}
	\bibinfo{author}{Ando, T., Fowler, A. B. \& Stern, F.}
    \newblock \bibinfo{title}{Electronic properties of two-dimensional systems}.
	\newblock \textit{\bibinfo{journal}{Rev. Mod. Phys.}} \textbf{\bibinfo{volume}{54}},
	\bibinfo{pages}{437} (\bibinfo{year}{1982}).
	
\bibitem{bychov_oscillatory_1984}
	\bibinfo{author}{ Bychov, Y. A. 
   \& Rashba, E.}
    \newblock \bibinfo{title}{Oscillatory effects and the magnetic susceptibility of carriers in inversion layers}.
	\newblock \textit{\bibinfo{journal}{J. Phys. C: Solid St. Phys.}} \textbf{\bibinfo{volume}{17}},
	\bibinfo{pages}{6039} (\bibinfo{year}{1984}).
	
\bibitem{datta_electronic_1990}
	\bibinfo{author}{Datta, S \& Das, B.}
    \newblock \bibinfo{title}{Electronic analog of the electro‐optic modulator}.
	\newblock \textit{\bibinfo{journal}{Appl. Phys. Lett.}} \textbf{\bibinfo{volume}{56}},
	\bibinfo{pages}{665} (\bibinfo{year}{1990}).
	
\bibitem{koga_rashba_2002}
	\bibinfo{author}{Koga, T., Nitta, J., Akazaki, T. \& Takayanagi, H.}
    \newblock \bibinfo{title}{Rashba Spin-Orbit Coupling Probed by the Weak Antilocalization Analysis in InAlAs/InGaAs/InAlAs Quantum Wells as a Function of Quantum Well Asymmetry}.
	\newblock \textit{\bibinfo{journal}{Phys. Rev Lett.}} \textbf{\bibinfo{volume}{89}},
	\bibinfo{pages}{046801} (\bibinfo{year}{2002}).


\bibitem{koo_control_2009}
	\bibinfo{author}{Koo, H. C., Kwon, J. H. Eom, J., Chang, J., Han, S. H. \& Johnson, M.}
    \newblock \bibinfo{title}{Control of Spin Precession in a Spin-Injected Field Effect Transistor}.
	\newblock \textit{\bibinfo{journal}{Science}} \textbf{\bibinfo{volume}{325}},
	\bibinfo{pages}{1515} (\bibinfo{year}{2009}).

\bibitem{yuan_high_2009}
	\bibinfo{author}{Yuan, H., Shimotani, H., Tsukazaki, A., Ohtomo, A., Kawasaki, M. \& Iwasa, Y.}
    \newblock \bibinfo{title}{High-Density Carrier Accumulation in ZnO Field‐Effect Transistors Gated by Electric Double Layers of Ionic Liquids.}.
	\newblock \textit{\bibinfo{journal}{Adv. Func. Mater.}} \textbf{\bibinfo{volume}{19}},
	\bibinfo{pages}{7} (\bibinfo{year}{2009}).
	
\bibitem{goldman_electrostatic_2014}
	\bibinfo{author}{Goldman, A. M.}
    \newblock \bibinfo{title}{Electrostatic Gating of Ultrathin Films}.
	\newblock \textit{\bibinfo{journal}{Annual Review of Materials Research}} \textbf{\bibinfo{volume}{44}},
	\bibinfo{pages}{45-63} (\bibinfo{year}{2014}).
	
\bibitem{ye_superconducting_2012}
	\bibinfo{author}{Ye, J. T., Zhang, Y. J., Akashi, R., Bahramy, M. S., Arita, R. \& Iwasa, Y.}
    \newblock \bibinfo{title}{Superconducting Dome in a Gate-Tuned Band Insulator}.
	\newblock \textit{\bibinfo{journal}{Science}} \textbf{\bibinfo{volume}{338}},
	\bibinfo{pages}{1193-1196} (\bibinfo{year}{2012}).
	
\bibitem{ueno_electric_2008}
	\bibinfo{author}{Ueno, K., Nakamura, S., Shimotani, H., Ohtomo, A., Kimura, N., Nojima, T., Aoki, H., Iwasa, Y. \& Kawasaki, M.}
    \newblock \bibinfo{title}{Electric-field-induced superconductivity in an insulator}.
	\newblock \textit{\bibinfo{journal}{Nat. Mater.}} \textbf{\bibinfo{volume}{7}},
	\bibinfo{pages}{855-858} (\bibinfo{year}{2008}).
	
\bibitem{ueno_discovery_2011}
	\bibinfo{author}{Ueno, K., Nakamura, S., Shimotani, H., Yuan, H. T., Kimura, N., Nojima, T., Aoki, H., Iwasa, Y. \& Kawasaki, M.}
    \newblock \bibinfo{title}{Discovery of superconductivity in KTaO$_3$ by electrostatic carrier doping}.
	\newblock \textit{\bibinfo{journal}{Nat. Nano.}} \textbf{\bibinfo{volume}{6}},
	\bibinfo{pages}{408-412} (\bibinfo{year}{2011}).
	
\bibitem{wen_evolution_2018}
	\bibinfo{author}{Wen, F., Cao, Y., Liu, X., Pal, B., Middey, S., Kareev, M. \& Cakhalian, J.}
    \newblock \bibinfo{title}{Evolution of ferromagnetism in two-dimensional electron gas of LaTiO$_3$/SrTiO$_3$}.
	\newblock \textit{\bibinfo{journal}{Appl. Phys. Lett.}} \textbf{\bibinfo{volume}{112}},
	\bibinfo{pages}{122405} (\bibinfo{year}{2018}).
	
\bibitem{yang_anisotropic_2015}
	\bibinfo{author}{Yang, Z., Kent, T. F., Yang, J., Jin, H., Heremans, J. P. \& Myers, R. C.}
    \newblock \bibinfo{title}{Anisotropic defect-induced ferromagnetism and transport in Gd-doped GaN two-dimensional electron gasses}.
	\newblock \textit{\bibinfo{journal}{Phys. Rev. B}} \textbf{\bibinfo{volume}{92}},
	\bibinfo{pages}{224416} (\bibinfo{year}{2015}).
	
\bibitem{wei_prediction_2017}
	\bibinfo{author}{Wei, L. Y., Lian, C. \& Meng, S.}
    \newblock \bibinfo{title}{Prediction of two-dimensional electron gas mediated magnetoelectric coupling at ferroelectric 
PbTiO$_3$/SrTiO$_3$ heterostructures}.
	\newblock \textit{\bibinfo{journal}{Phys. Rev. B}} \textbf{\bibinfo{volume}{95}},
	\bibinfo{pages}{184102} (\bibinfo{year}{2017}).
	
\bibitem{yamada_electrically_2011}
	\bibinfo{author}{Yamada, Y., Ueno, K., Fukumura, T., Yuan, H. T., Shimotani, H., Iwasa, Y., Gu, L., Tsukimoto, S., Ikuhara, Y. \& Kawasaki, M.}
    \newblock \bibinfo{title}{Electrically Induced Ferromagnetism at Room Temperature in Cobalt-Doped Titanium Dioxide}.
	\newblock \textit{\bibinfo{journal}{Science}} \textbf{\bibinfo{volume}{332}},
	\bibinfo{pages}{1065-1067} (\bibinfo{year}{2011}).
	
\bibitem{chhowalla_chemistry_2013}
	\bibinfo{author}{Chhowalla, M., Shin, H. S., Eda, G, Li, L.-J.,  Loh, K. P. \& Zhang, H.}
    \newblock \bibinfo{title}{The chemistry of two-dimensional layered transition metal dichalcogenide nanosheets}.
	\newblock \textit{\bibinfo{journal}{Nat. Chem.}} \textbf{\bibinfo{volume}{5}},
	\bibinfo{pages}{263-275} (\bibinfo{year}{2013}).
	

	
\bibitem{rossnagal_origin_2011}
\bibinfo{author}{Rossnagal, K.}
    \newblock \bibinfo{title}{On the origin of charge-density waves in select layered transition-metal dichalcogenides}.
\newblock \textit{\bibinfo{journal}{J. Phys. Condens. Matter}} \textbf{\bibinfo{volume}{23}},
\bibinfo{pages}{21} (\bibinfo{year}{2011}).

\bibitem{xu_spin_2014}
\bibinfo{author}{Xu, X., Yao, W., Xiao, D. \& Heinz, T. F.}
    \newblock \bibinfo{title}{Spin and pseudospins in layered transition metal dichalcogenides}.
\newblock \textit{\bibinfo{journal}{Nat. Phys.}} \textbf{\bibinfo{volume}{10}},
\bibinfo{pages}{343-350} (\bibinfo{year}{2014}).
    
\bibitem{manzeli_2D_2017}
\bibinfo{author}{Manzeli, S., Ovchinnikov, D., Pasquier, D., Yazyev, O. V. \& Kis, A.}
    \newblock \bibinfo{title}{2D transition metal dichalcogenides}.
\newblock \textit{\bibinfo{journal}{Nat. Rev. Mater.}} \textbf{\bibinfo{volume}{2}},
\bibinfo{pages}{17033} (\bibinfo{year}{2014}).

\bibitem{bahramy_ubiquitous_2018}
	\bibinfo{author}{Bahramy, M. S., Clark, O. J., Yang, B.-J., Feng, J., Bawden, L., Riley, J. M., Markovi\'c, I., Mazzola, F., Sunko, V., Biswas, D., Cooil, S. P., Jorge, M., Wells, J. W., Leandersson, M., Balasubramanian, T., Fujii, J., Vobornik, I., Rault, J. E., Kim, T. K., Hoesch, M., Okawa, K., Asakawa, M., Sasagawa, T., Eknapakul, T., Meevasana, W. \& King, P. D. C.}
    \newblock \bibinfo{title}{Ubiquitous formation of bulk dirac cones and topological surface states from a single orbital manifold in transition-metal dichalcogenides}.
	\newblock \textit{\bibinfo{journal}{Nat. Mater.}} \textbf{\bibinfo{volume}{17}},
	\bibinfo{pages}{23-27} (\bibinfo{year}{2018}).
	
\bibitem{amani_electrical_2013}
	\bibinfo{author}{Amani, M., Chin, M. L., Birdwell, A. G., O'Regan, T. P., Najmaei, S., Liu, Z., Ajayan, P. M., Lou, J. \& Dubey, M.}
    \newblock \bibinfo{title}{Electrical performance of monolayer MoS$_2$ field-effect transistors prepared by chemical vapor deposition}.
	\newblock \textit{\bibinfo{journal}{Appl. Phys. Lett.}} \textbf{\bibinfo{volume}{102}},
	\bibinfo{pages}{193107} (\bibinfo{year}{2013}).
	

\bibitem{gong_electric_2014}
	\bibinfo{author}{Gong, K., Zhang, L., Liu, D, Liu, K., Zhu, Y., Zhao, Y. \& Guo, H.}
    \newblock \bibinfo{title}{Electric control of spin in monolayer WSe$_2$ field effect transistors}.
	\newblock \textit{\bibinfo{journal}{Nanotechnology}} \textbf{\bibinfo{volume}{25}},
	\bibinfo{pages}{43} (\bibinfo{year}{2014}).
	
	 
\bibitem{mak_valley_2014}
	\bibinfo{author}{Mak, K. F., McGill, K. L., Park, J. \& McEuen, P. L.}
    \newblock \bibinfo{title}{The valley Hall effect in MoS$_2$ transistors}.
	\newblock \textit{\bibinfo{journal}{Science}} \textbf{\bibinfo{volume}{344}},
	\bibinfo{pages}{1489-1492} (\bibinfo{year}{2014}).
	
\bibitem{riley_negative_2015}
	\bibinfo{author}{Riley, J. M., Meevasana, W., Bawden, L., Asakawa, M., Takayama, T., Eknapakul, T., Kim, T. K., Hoesch, M., Mo, S.-K., Takagi, H., Sasagawa, T., Bahramy, M. S. \& King, P. D. C.}
    \newblock \bibinfo{title}{Negative electronic compressibility and tunable spin splitting in WSe$_2$}.
	\newblock \textit{\bibinfo{journal}{Nat. Nano.}} \textbf{\bibinfo{volume}{10}},
	\bibinfo{pages}{1043-1047} (\bibinfo{year}{2015}).
	
\bibitem{yuan_zeeman_2013}
	\bibinfo{author}{Yuan, H., Bahramy, M. S., Morimoto, K., Wu, S., Nomura, K., Yang, B.-J., Shimotani, H., Suzuki, R., Toh, M., Kloc, C., Xu, X., Arita, R., Nagosa, N. \& Iwasa, Y.}
    \newblock \bibinfo{title}{Zeeman-type spin splitting controlled by an electric field}.
	\newblock \textit{\bibinfo{journal}{Nat. Phys.}} \textbf{\bibinfo{volume}{9}},
	\bibinfo{pages}{563-569} (\bibinfo{year}{2013}).
	
\bibitem{li_controlling_2016}
	\bibinfo{author}{Li, L. J., O'Farrell, E. C. T., Loh, K. P., Eda, G., \"Ozyilmaz, B. \& Castro Neto, A. H.}
    \newblock \bibinfo{title}{Controlling many-body states by the electric-field effect in a two-dimensional material}.
	\newblock \textit{\bibinfo{journal}{Nature}} \textbf{\bibinfo{volume}{534}},
	\bibinfo{pages}{21-22} (\bibinfo{year}{2016}).
	
\bibitem{lu_evidence_2015}
	\bibinfo{author}{Lu, J. M., Zheliuk, O., Leermakers, I., Yuan, N. F. Q., Law, K. T. \& Ye, J. T.}
    \newblock \bibinfo{title}{Evidence for two-dimensional Ising superconductivity in gated MoS$_2$}.
	\newblock \textit{\bibinfo{journal}{Science}} \textbf{\bibinfo{volume}{350}},
	\bibinfo{pages}{6266, 1353-1357} (\bibinfo{year}{2015}).
	
\bibitem{saito_superconductivity_2016}
	\bibinfo{author}{Saito, Y., Nakamura, Y., Bahramy, M. S., Kohama, Y., Ye, J., Kasahara, Y., Nakagawa, Y., Onga, M., Tokunaga, M., Nojima, T., Yanase, Y. \& Iwasa, Y.}
    \newblock \bibinfo{title}{Superconductivity protected by spin-valley locking in ion-gated MoS$_2$}.
	\newblock \textit{\bibinfo{journal}{Nat. Phys.}} \textbf{\bibinfo{volume}{12}},
	\bibinfo{pages}{144-149} (\bibinfo{year}{2016}).
	
\bibitem{yang_quantum_2018}
	\bibinfo{author}{Yang, H., Schmidt, M., Suss, V., Chan, M., Balakirev, F. F., McDonald, R. D., Parkin, S. S. P., Felser, C., Yan, B. \& Moll, P. J. W.}
    \newblock \bibinfo{title}{Quantum oscillations in the type-II Dirac semimetal candidate PtSe$_2$}.
	\newblock \textit{\bibinfo{journal}{New J. Phys.}} \textbf{\bibinfo{volume}{20}},
	\bibinfo{pages}{043008} (\bibinfo{year}{2018}).

\bibitem{hoesch_facility_2017}
	\bibinfo{author}{Hoesch, M., Kim, T. K., Dudin, P., Wang, H., Scott, S., Harris, P., Patael, S., Matthews, M., Hawkins, D., Alcock, S. G., Richter, T., Mudd, J. J., Basham, M., Pratt, L., Leicester, P., Longhi, E. C., Tamai, A. \& Baumberger, F.} 
    \newblock \bibinfo{title}{A facility for the analysis of the
electronic structures of solids and their surfaces by synchrotron
radiation photoelectron spectroscopy}.
	\newblock \textit{\bibinfo{journal}{Rev. Sci.
Instr.}} \textbf{\bibinfo{volume}{88}},
	\bibinfo{pages}{013106} (\bibinfo{year}{2017}).

\bibitem{Eknapakul}
    \bibinfo{author}{Eknapakul, T.} \textit{et~al.}
\newblock \bibinfo{title}{Nearly-free-electron system of monolayer Na on the surface of single-crystal ${\mathrm{HfSe}}_{2}$}.
	\newblock \textit{\bibinfo{journal}{Phys. Rev. B.}} \textbf{\bibinfo{volume}{94}},
	\bibinfo{pages}{201121} (\bibinfo{year}{2016}).	


\bibitem{wien}
	\bibinfo{author}{Balaha, P.} \textit{et~al.}
	\newblock \bibinfo{title}{{\sc wien2k} package, Version 13.1}.
	(\bibinfo{year}{2013}).


\bibitem{souza}
	\bibinfo{author}{Souza, I., Marzari, N. \& Vabderbilt, D.}
	\newblock \bibinfo{title}{Maximally localized Wannier functions for entangled energy bands}.
	\newblock \textit{\bibinfo{journal}{Phys. Rev. B.}} \textbf{\bibinfo{volume}{65}},
	\bibinfo{pages}{035109} (\bibinfo{year}{2001}).	

\bibitem{mostofi}
	\bibinfo{author}{Mostofi, A. A., Yates, J. R., Lee, Y.-S., Souza, I., Vanderbilt, D. \& Marzari, N.} 
	\newblock \bibinfo{title}{Wannier90: a tool for obtaining maximally localized Wannier functions}.
	\newblock \textit{\bibinfo{journal}{Comp. Phys. Commun.}} \textbf{\bibinfo{volume}{178}},
	\bibinfo{pages}{685-699} (\bibinfo{year}{2008}).	
\bibitem{kunes}
	\bibinfo{author}{Kunes, J.} \textit{et~al.}
	\newblock \bibinfo{title}{WIEN2WANNIER: from linearized augmented plane waves to maximally localized Wannier functions}.
	\newblock \textit{\bibinfo{journal}{Comp. Phys. Commun.}} \textbf{\bibinfo{volume}{181}},
	\bibinfo{pages}{1888-1895} (\bibinfo{year}{2010}).

\bibitem{lei_2017}
	\bibinfo{author}{Lei, J.-Q., Ke, L., Sha, H., \&  Zhou, X.,-L.}
    \newblock \bibinfo{title}{The comparative study on bulk-PtSe$_2$ and 2D 1-Layer-PtSe$_2$ under high pressure via first-principle calculations}.
	\newblock \textit{\bibinfo{journal}{Theor. Chem. Acc.}} \textbf{\bibinfo{volume}{136}},
	\bibinfo{pages}{97} (\bibinfo{year}{2017}).
	
\bibitem{guo_electronic_1986}
	\bibinfo{author}{Guo, G. Y. \& Liang, W. Y.}
    \newblock \bibinfo{title}{The electronic structures of platinum dichalcogenides: PtS$_2$, PtSe$_2$ and PtTe$_2$}.
	\newblock \textit{\bibinfo{journal}{J. Phys. C: Solid State Phys.}} \textbf{\bibinfo{volume}{19}},
	\bibinfo{pages}{995} (\bibinfo{year}{1986}).



\bibitem{huang_type_2016}
\bibinfo{author}{Huang, H., Zhou, S. \& Duan, W.}
    \newblock \bibinfo{title}{Type-II Dirac Fermions in the Transition Metal Dichalcogenide PtSe$_2$ Class}.
\newblock \textit{\bibinfo{journal}{Phys. Rev. B}} \textbf{\bibinfo{volume}{94}},
\bibinfo{pages}{121117} (\bibinfo{year}{2016}).


\bibitem{zhang_experimental_2017}
\bibinfo{author}{Zhang, K., Yan. M., Zhang, H., Huang, H., Arita, M., Sun, Z., Duan, W., Wu, Y. \& Zhou, S.}
    \newblock \bibinfo{title}{Experimental evidence of type-II Dirac fermions in PtSe$_2$}.
\newblock \textit{\bibinfo{journal}{Phys. Rev. B}} \textbf{\bibinfo{volume}{96}},
\bibinfo{pages}{125102} (\bibinfo{year}{2017}).


\bibitem{yan_lorentz_2017}
\bibinfo{author}{Yan, M., Huang, H., Zhang, K., Wang, E., Yao, W., Deng, K., Wan, G., Zhang, H., Arita, M., Yang, H., Sun, Z., Yao, H., Wu, Y., Fan, S., Duan, W. \& Zhou, S.}
    \newblock \bibinfo{title}{Lorentz-violating type-II Dirac fermions in transition metal dichalcogenide PtTe$_2$}.
\newblock \textit{\bibinfo{journal}{Nat. Commun.}} \textbf{\bibinfo{volume}{8}},
\bibinfo{pages}{257} (\bibinfo{year}{2017}).



	
\bibitem{Supp}
    \newblock \bibinfo{title}{see Supplemental Information}.

	
\bibitem{clark_fermiology_2018}
\bibinfo{author}{Clark, O. J., Neat, M. J.,  Okawa, K., Bawden, L., Markovi\'c, I., Mazzola, F., Feng, J., Sunko, V., Riley, J. M., Meevasana, W., Fujii, J., Vobornik, I., Kim, T. K., Hoesch, M., Sasagawa, T., Bahramy, Wahl, P., M. S. \& King, P. D. C.}
    \newblock \bibinfo{title}{Fermiology and Superconductivity of Topological Surface States in PdTe$_2$}.
\newblock \textit{\bibinfo{journal}{Phys. Rev. Lett.}} \textbf{\bibinfo{volume}{120}},
\bibinfo{pages}{156401} (\bibinfo{year}{2018}).

\bibitem{Riley_NP}
	\bibinfo{author}{Riley, J.M., Mazzola, F., Dendzik, M., Michiardi, M., Takayama, T., Bawden, L., Granerod, C., Leandersson, M., Balasubramanian, T., Hoesch, M.,  Kim, T. K., Takagi, H., Meevasna, W., Hofmann, Ph., Bahramy, M.S., Wells, J., \& King, P. D. C.}
    \newblock \bibinfo{title}{Direct observation of spin-polarized bulk bands in an inversion-symmetric semiconductor}.
	\newblock \textit{\bibinfo{journal}{Nature Phys.}} \textbf{\bibinfo{volume}{10}},
	\bibinfo{pages}{835-839} (\bibinfo{year}{2014}).

\bibitem{Kim_NL}
	\bibinfo{author}{Kang, M., Kim, B., Ryu, S.H., Jung, S.W., Kim, J., Moreschini, L., Jozwiak, C., Rotenberg, E., Bostwick, A., \& Kim, K.S.}
    \newblock \bibinfo{title}{Universal Mechanism of Band-Gap Engineering in Transition-Metal Dichalcogenides}.
	\newblock \textit{\bibinfo{journal}{Nano Lett.}} \textbf{\bibinfo{volume}{17}},
	\bibinfo{pages}{1610-1615} (\bibinfo{year}{2017}).
	
\bibitem{Zhang_NP}
	\bibinfo{author}{Zhang, X., Liu, Q., Luo, J.-W., Freeman, A.J., \& Zunger, A.}
    \newblock \bibinfo{title}{Hidden spin polarization in inversion-symmetric bulk crystals}.
	\newblock \textit{\bibinfo{journal}{Nature Phys.}} \textbf{\bibinfo{volume}{10}},
	\bibinfo{pages}{387-393} (\bibinfo{year}{2014}).	
	
 \bibitem{sunko_maximal_2017}
	\bibinfo{author}{Sunko, V., Rosner, H., Kushwaha, P., Khim, S., Mazzola, F., Bawden, L., Clark, O. J., Riley, J. M., Kasinathan, D., Haverkort, M. W., Kim, T. K., Hoesch, M., Fujii, J., Vobornik, I., Mackenzie, A. P. \& King, P. D. C.}
    \newblock \bibinfo{title}{Maximal Rashba-like spin splitting via kinetic-energy-coupled inversion-symmetry breaking}.
	\newblock \textit{\bibinfo{journal}{Nature}} \textbf{\bibinfo{volume}{549}},
	\bibinfo{pages}{492-496} (\bibinfo{year}{2017}).



   
	
	\bibitem{park_orbital_2011}
	\bibinfo{author}{Park, S. R., Kim, C. H., Yu, J., Han, J. H., \& Kim, C.}
    \newblock \bibinfo{title}{Orbital-Angular-Momentum Based Origin of Rashba-Type Surface Band Splitting}.
	\newblock \textit{\bibinfo{journal}{Phys. Rev. Lett.}} \textbf{\bibinfo{volume}{107}},
	\bibinfo{pages}{156803} (\bibinfo{year}{2011}).


  \bibitem{kim_nature_2014}
	\bibinfo{author}{Kim, P., Kang, K. T., Go, G. \& Han, J. H.}
    \newblock \bibinfo{title}{Nature of orbital and spin Rashba coupling in the surface bands of SrTiO$_3$ and KTaO$_3$}.
	\newblock \textit{\bibinfo{journal}{Phys. Rev. B}} \textbf{\bibinfo{volume}{90}},
	\bibinfo{pages}{205423} (\bibinfo{year}{2014}).
	

\bibitem{moench_semiconductor}
	\bibinfo{author}{Moench, W.} 
    \newblock \bibinfo{title}{Semiconductor Surfaces and Interfaces}.
 (\bibinfo{year}{Springer, 2001}).
    
\bibitem{metadata}
    \newblock \bibinfo{title}{https://doi.org/10.17630/803c15f3-fc3d-4a80-a99c-c2154dce358b}.
    



\end{thebibliography}
\end{document}